\documentclass[%
 reprint,
 amsmath,amssymb,prl,aps]{revtex4-1}
\usepackage{amsmath}    
\usepackage{graphicx}   
\usepackage{verbatim}   
\usepackage{color}      
\usepackage{multirow,tabularx}
\usepackage{hyperref}   
\begin{document}	

\title{Spin-orbit coupling and spin relaxation in phosphorene: Intrinsic versus extrinsic effects}
\author{Marcin Kurpas, Martin Gmitra, and Jaroslav Fabian}
\affiliation{Institute for Theoretical Physics, University of Regensburg,\\93040 Regensburg, Germany}

\date{\today}

\begin{abstract}
First-principles calculations of the essential spin-orbit and spin relaxation properties of phosphorene are performed. Intrinsic spin-orbit coupling induces spin mixing with the
probability of $b^2 \approx 10^{-4}$, exhibiting a large anisotropy, following the anisotropic
crystalline structure of phosphorene. For realistic values of the momentum relaxation times, 
the intrinsic (Elliott--Yafet) spin relaxation times are hundreds of picoseconds to nanoseconds.
Applying a transverse electric field (simulating gating and substrates) generates extrinsic $C_{2v}$ symmetric spin-orbit fields
in phosphorene, which activate the D'yakonov--Perel' mechanism for spin relaxation. It 
is shown that this extrinsic spin relaxation also has a strong anisotropy, and can dominate
over the Elliott-Yafet one for strong enough electric fields. Phosphorene on substrates can 
thus exhibit an interesting interplay of both spin relaxation mechanisms, whose individual roles could be deciphered using our results.

\end{abstract}

\maketitle


Phosphorene is a monolayer of black phosphorus \cite{brown_refinement_1965,cartz_effect_1979,%
keyes_electrical_1953,maruyama_synthesis_1981,narita_electrical_1983}, 
exhibiting a direct band gap of 2~eV \cite{castellanos-gomez_isolation_2014, liang_electronic_2014} and large anisotropic mobility 
\cite{castellanos-gomez_isolation_2014, liu_phosphorene_2014,qiao_high-mobility_2014}. Unlike graphene,
phosphorene is a semiconductor, and unlike two-dimensional transition-metal dichalcogenides, which are semiconductors too, phosphorene is distinctly anisotropic
thanks to its puckered atomic structure. The semiconductor property makes phosphorene
suitable for electronic \cite{li_black_2014} and spintronics applications \cite{Zutic2004:RMP, Fabian2007:APS}, in particular for bipolar spin diodes and 
transistors \cite{zutic_bipolar_2006}, while the anisotropy enables directional control of the essential spin properties, such as spin-orbit coupling and spin relaxation. In contrast to 
graphene, whose spin properties are by now well established \cite{han_graphene_2014},
there is no unified 
picture of the spin-orbit coupling and spin relaxation in phosphorene.

\begin{figure}[b] 
\centering
\includegraphics[width=0.99\columnwidth]{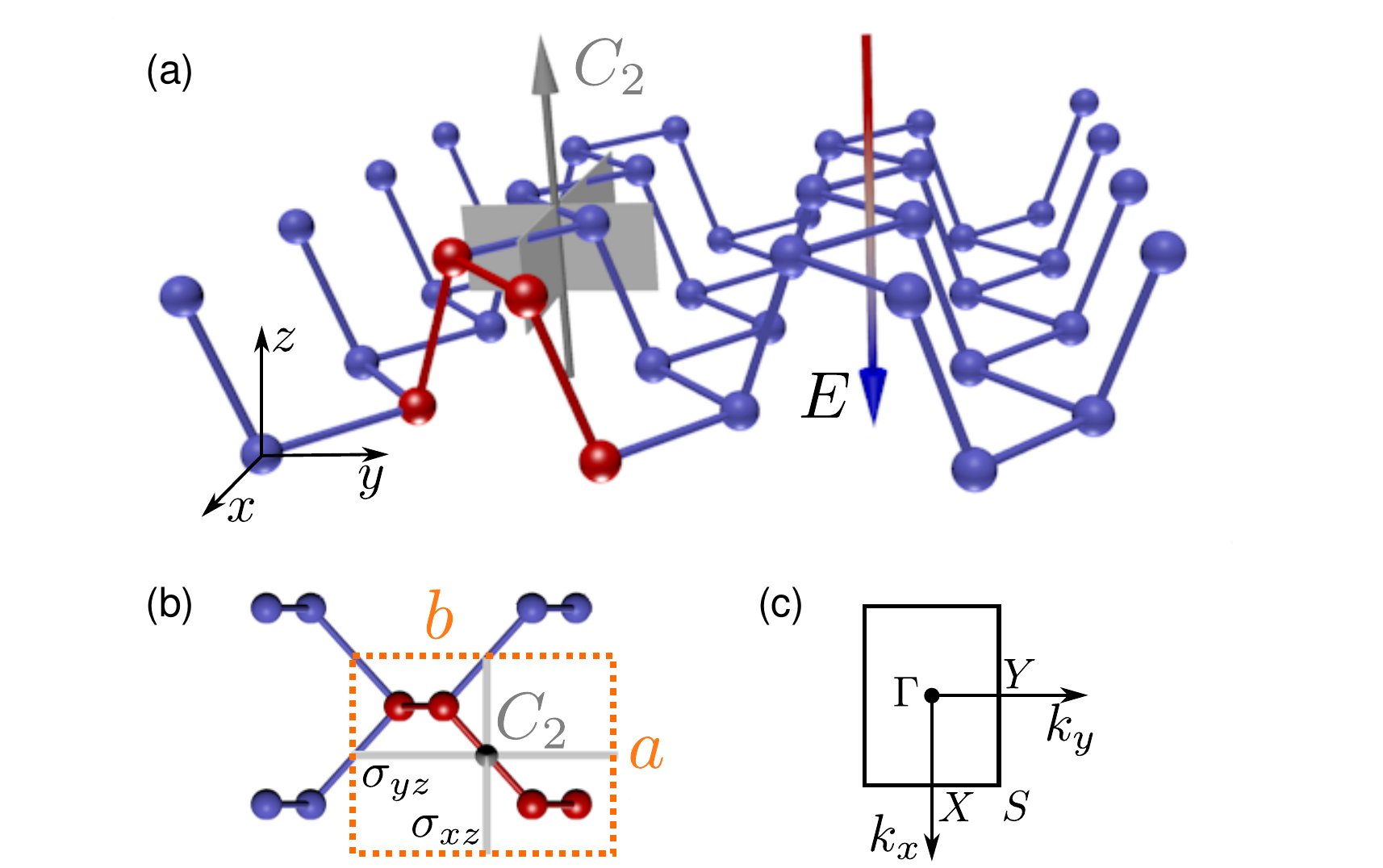}
\caption{\label{fig:bp3d} (Color online) Essence of phosphorene crystal structure. 
(a)~Schematic of a single layer of black phosphorus with drawn symmetry axis $C_2$, 
mirror planes $\sigma_{xz}$ and $\sigma_{yz}$ and the vector of electric field $E$.
The red colored atoms form the unit cell of phosphorene, the gradient of color 
of the electric field vector corresponds to higher (red) and lower (blue) 
electrostatic potential. 
(b)~Top view to the structure. The unit cell is marked by orange dashed line. 
(c)~The first Brillouin zone of phosphorene with labels of high symmetry points.}
\end{figure}

Phosphorene can be extracted from black phosphorus by mechanical \citep{xia_rediscovering_2014,liu_phosphorene_2014,li_black_2014} 
or liquid \citep{brent_liquid_exf,kang_solvent_2015} cleavage techniques.
Inside phosphorene layers, each phosphorus atom is covalently bonded with three adjacent 
phosphorus atoms to form a puckered honeycomb structure due to $sp^3$ hybridization, 
see Fig.~\ref{fig:bp3d}. The puckered structure can be viewed as a two-layer system in which
the bonding energy is dominated by the in-plane bonds ($pp\sigma$ and $pp\pi$) that 
are much stronger than the bonds connecting the two sublayers ($pp\pi$) \citep{li_electrons_2014}. 
Similarly to graphene, the edges of phosphorene form zig-zag (along $x$ axis) and armchair 
(along $y$ axis) chains [see Fig. \ref{fig:bp3d}b)]. 

Black phosphorus is described by 
the nonsymmorphic $D_{2h}$ point group being isomorphic with the $C_{mca}$ space group.
Phosphorene shares the same point group symmetry as its bulk counterpart. 
Both structures have inversion symmetry leading to spin degenerate eigenstates. Spin-orbit 
coupling leads to the spin-mixing of the Pauli spinors---the {\it intrinsic effect}.   
When inversion symmetry of phosphorene is broken by an applied transverse electric
field $\mathbf{E}$ or a substrate, the point group is reduced to 
nonsymmorphic $C_{2v}$, with the principal $C_2$ axis parallel to the direction of the 
electric field and two mirror planes $\sigma_{xz}$ and $\sigma_{yz}$ [Fig. \ref{fig:bp3d} (a,b)]. In this case the spin degeneracy is lifted---the {\it extrinsic} (Rashba) effect.

%

From the spintronics perspective two questions are particularly important to address: 
(i)~what is the intrinsic and extrinsic spin-orbit coupling (SOC) in phosphorene and
(ii)~what are the relevant spin relaxation time scales. 
The answer to the first question has been partially answered by Popovi\'{c} et al. \cite{Popovic_2015} who showed that the extrinsic Rashba effect, due to external electric fields, is anisotropic with respect to the two principal directions in the crystal. 
The second question has been addressed within $k \cdot p$ theory \cite{li_electrons_2014} 
for the intrinsic effects only.

Here we employ first-principles calculations to address both questions, providing state-of-the-art most
realistic results for the extrinsic and intrinsic effects. 
First, we find that intrinsic SOC lifts degeneracy of the valence and conduction bands at the S point, by splitting the bands of about 17.5~meV and 14~meV, respectively. 
The extrinsic Rashba SOC is much weaker, of the order of tens of $\mu$eV close to the $\Gamma$ point (for electric fields of 1 V/nm), and is found to be significantly anisotropic for the valence band only.
Second, we predict the spin lifetime in phosphorene to be hundreds of picoseconds up to nanoseconds, for the experimentally relevant mobilities. We find that for no and small electric fields up to ($\text{E} \le 2.5$~V/nm) and carrier densities up to $8\cdot 10^{12}$~cm$^{-2}$, the dominant spin relaxation mechanism is the Elliott--Yafet \cite{elliott_theory_1954,Yafet_1963}.
For the in-plane spin orientation the relaxation is almost twice the slower than for spins oriented out-of-plane.
By increased transverse electric field the D'yakonov--Perel' \cite{dyakonov_1971R} mechanism starts to be the most effective. For carrier density $n
\approx 3 \cdot 10^{12}$~cm$^{-2}$ it overtakes the Elliott--Yafet's at $E=4$~V/nm for holes and $E=5$~V/nm for electrons. 
As a result, the substrates can be essential for spin dynamics in phosphorene due to an interplay between the D'yakonov--Perel' and the Elliott--Yafet relaxation mechanisms.

Intrinsic phosphorene belongs to the family of centrosymmetric crystals for which the dominant spin relaxation mechanism is the Elliott--Yafet scattering \cite{elliott_theory_1954,Yafet_1963}. The strength of the scattering can be quantified by the spin admixture coefficient $b^2$ \cite{elliott_theory_1954}, which is a signature
of intrinsic spin-orbit coupling.
The time-reversal and space-inversion symmetries require two Bloch states~$\Psi_{n,k}^{\uparrow}(\mathbf{r})$ and $\Psi_{n,k}^{\downarrow}(\mathbf{r})$ of the same band $n$ and momentum $k$ to be degenerate (Kramer's doublets). 
Due to spin-orbit coupling these states are mixtures of spin up $\vert\uparrow\rangle$ and spin down 
$|\downarrow\rangle$ Pauli spinors: $\Psi_{n,\mathbf{k}}^{\uparrow}(\mathbf{r})= \left[a_{n,\mathbf{k}}(\mathbf{r})|\uparrow\rangle + b_{n,\mathbf{k}}(\mathbf{r})|\downarrow\rangle \right] e^{i\mathbf{k}\mathbf{r}}$, $\Psi_{n,\mathbf{k}}^{\downarrow}(\mathbf{r})= [a_{n,\mathbf{-k}}^{\ast}(\mathbf{r})|\downarrow\rangle - b_{n,\mathbf{-k}}^{\ast}(\mathbf{r})|\uparrow\rangle ] e^{i\mathbf{k}\mathbf{r}}$. For a generic Bloch state 
the modulation functions $a_{n,\mathbf{k}}(\mathbf{r})$ and $b_{n,\mathbf{k}}(\mathbf{r})$ are selected 
to diagonalize the spin magnetic moment along the chosen direction, corresponding to the injected spin in experiment;
for weak spin-orbit coupling, $b_{n,\mathbf{k}}(\mathbf{r})$ stands for the small spin component being admixed to the large spin component $a_{n,\mathbf{k}}(\mathbf{r})$, i. e., $\vert a_{n,\mathbf{k}}(\mathbf{r})\vert^2 \gg \vert b_{n,\mathbf{k}}(\mathbf{r})\vert^2$. 
The Eliott--Yafet scattering parameter $b^2$ is defined as the Fermi surface average of the unit cell integrated admixture coefficient $b^2_{n,\mathbf{k}}$, 
\begin{eqnarray}
b^2 &=&\langle b^2_{n,\mathbf{k}} \rangle=[\rho(\varepsilon_{\rm F}) S_{\rm BZ}]^{-1}\int_{\rm FS} b^2_{n,\mathbf{k}}/\vert \hbar v_{\rm F}(\varepsilon_{\rm F})\vert dk, \qquad \\ 
b^2_{n,\mathbf{k}} &=& \int \vert b_{n,\mathbf{k}}(\mathbf{r})\vert^2 d^3r, 
\label{eq_average}
\end{eqnarray}
where $0\leq b^2 \leq0.5$, $\rho(\varepsilon_{\rm F})$ 
is the density of states per spin at the Fermi level, $v_{\rm F}$ is the Fermi velocity, and $S_{\rm BZ}$ is the area of the Brillouin zone.
If the scattering potential is spin-independent (scalar impurities and phonons), 
the intrinsic SOC leads to spin-flip scattering. The Elliott--Yafet mechanism gives for the spin relaxation rate, 
\cite{elliott_theory_1954,fabian_spin_1998}
\begin{equation}
\tau_{\rm s,EY}^{-1} \approx 4 b^2/\tau_p^{-1},
\label{eq_ey}
\end{equation}
where $\tau_p^{-1}$ is the momentum relaxation rate.

Extrinsic effects appear once the space inversion symmetry gets broken, e.g., by a substrate or external fields. The spin degeneracy gets lifted and another spin relaxation mechanism appears: D'yakonov--Perel' \cite{dyakonov_1971R}. 
This mechanism can be viewed as a motional narrowing of the spin precession in a fluctuating (due to momentum scattering) emerging spin-orbit field $\mathbf{\Omega}_\mathbf{k}$, which is related to the spin splitting as
\begin{equation} \label{eq_ss}
H_{\rm ex} = \frac{\hbar}{2} \mathbf{\Omega}_{\bf k} \cdot \boldsymbol{\sigma},
\end{equation}
where $\boldsymbol{\sigma}$ is the vector of Pauli matrices.
In the relevant limit of small correlation times ($\Omega \tau_p \ll 1$),  i. e., when the precession angle between the scattering events is small, the spin relaxation rate becomes, \cite{dyakonov_1971R}
\begin{equation}
\tau_{\rm s,DP}^{-1}=\Omega_{\perp}^2\tau_p,
\label{eq_dp}
\end{equation}
where $\Omega_{\perp}^2=\langle\Omega^2_{\mathbf{k}, \perp}\rangle$ is the Fermi contour average of the 
squared spin-orbit field projected to the plane perpendicular to the spin orientation.

\begin{figure}
\centering
\includegraphics[width=0.99\columnwidth]{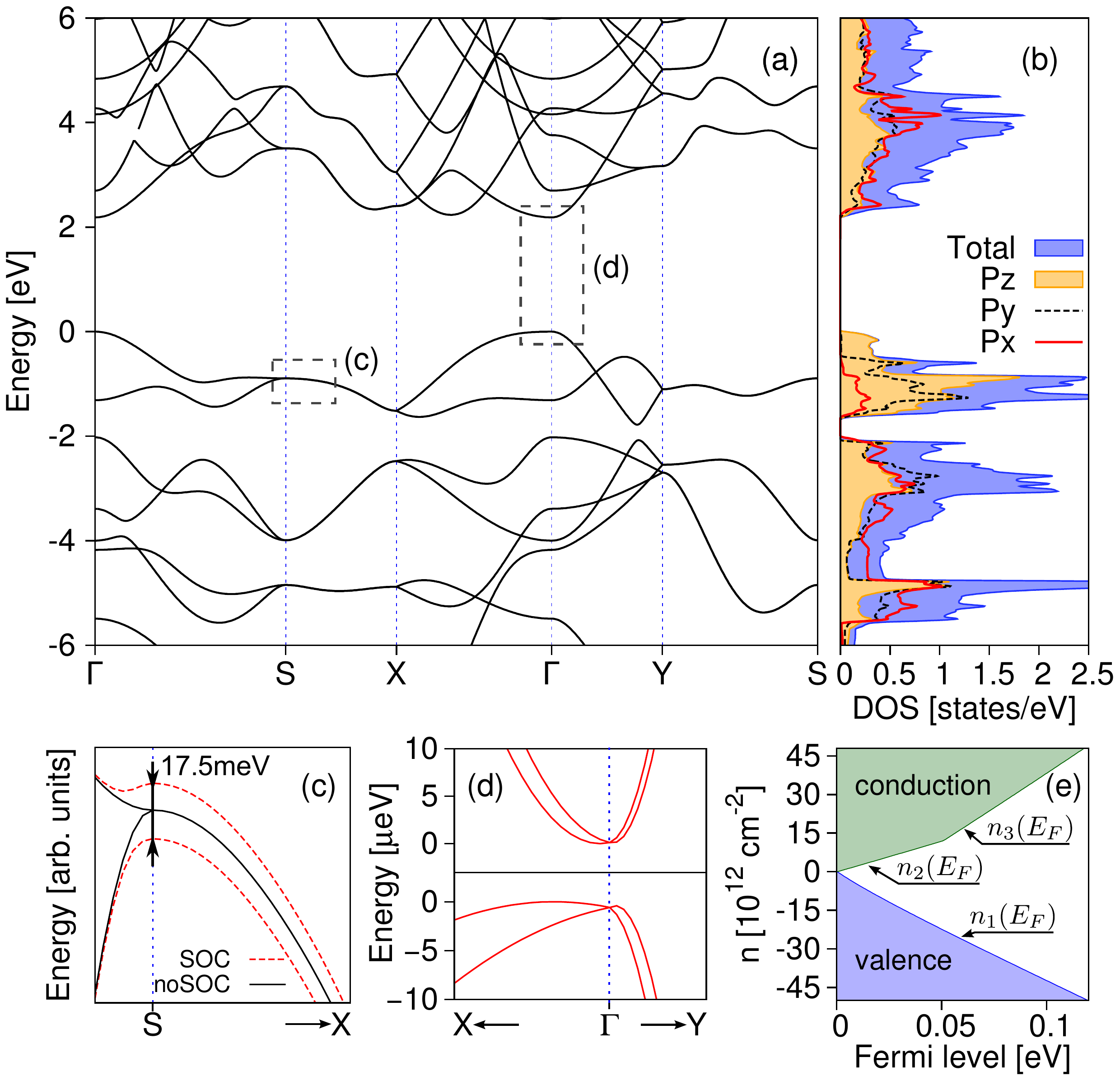}
\caption{\label{fig:bs_dos} (Color online) Calculated electronic properties of phosphorene
using LDA+mBJ exchange-correlation functional.
(a)~Band structure along high symmetry lines. The marked areas (c) and (d) are zoomed in the bottom row of the figure.
(b)~$p$-orbital resolved (yellow-filled curve, solid and dashed lines) and total (blue-filled curve) density of states.
(c) ~Splitting of the valence band along the S--X line due to intrinsic spin-orbit coupling. (d) Sketch of the 
extrinsic (Rashba) spin-orbit coupling effect to the band structure close to $\Gamma$. 
 (e)~Carrier concentration as a function of the Fermi level. The value $n$=0 corresponds to the Fermi level at the valence (conduction) band maximum (minimum). Positive values of $n$ correspond to electron, negative to hole doping. 
Empirical fits to the first-principles results give,
$n_1(\varepsilon_{\rm F})=-243.3\varepsilon_{\rm F}^2-441.4\varepsilon_{\rm F}+0.86$, $n_2(\varepsilon_{\rm F})=238.9\varepsilon_{\rm F}-0.14$, $n_3(\varepsilon_{\rm F})=537.9\varepsilon_{\rm F}-15.52$, where $\varepsilon_{\rm F}$ is the Fermi level in eV and density $n$ in $10^{12}$~cm$^{-2}$.
}
\end{figure}

In our first-principles calculations we used the initial crystal structure parameters from Ref.~\cite{brown_refinement_1965} 
for bulk black phosphorus. A sheet of phosphorene was placed in vacuum of {20~\AA} and fully relaxed using 
quasi--Newton variable--cell scheme as implemented in the {\sc Quantum Espresso} \citep{QE-2009} package.
Positions of atoms have been relaxed in all directions with the force convergence threshold 
$10^{-4}$~Ry/a.u. and total energy convergence condition $10^{-5}$~Ry/a.u..
The norm-conserving pseudopotential, with kinetic energy cutoffs of 70~Ry and 280~Ry for the 
wavefunction and charge density respectively, has been used along with the PBE exchange-correlation functional
\cite{Pedrew_1996}. 
Obtained structural parameters are summarized in Ref.~\cite{SM}.\\
Further electronic structure calculations have been performed using the full-potential linearized augmented plane-wave
method as implemented in all-electron code package {\sc Wien2k} \cite{wien2k}. Self-consistency has been achieved 
for $16\times 12\times 1$ Monkhorst-Pack $k$-point grid with 151 $k$--points in the irreducible wedge of the 
Brillouin zone. SOC has been included fully relativistically for core electrons while five valence electrons
have been treated within second variational step method \cite{Singh2006:book}.
For the calculations with the transverse electric field we considered vacuum size of 25~\AA.
It is known that standard DFT methods underestimate the bandgap of semiconductors. Theoretical bandgaps of phosphorene spread  between $0.7$~eV and $2.2$~eV depending on the method of calculations\cite{rodin_strain_induced_2014,qiao_high-mobility_2014,liu_phosphorene_2014,hu_mechanical_2014,rudenko_quasiparticle_2014,tran_layer-controlled_2014}. 
On the other hand, recent experimental reports suggest the bandgap of phosphorene of about 2~eV \cite{liang_electronic_2014,wang_highly_2015}.
It has been reported for standard semiconductors \cite{Chantis2006:PRL} that the underestimation of the bandgap impairs significantly SOC effects.
Therefore, to consider a realistic bandgap we perform the calculations with
undressed LDA functional along with the modified Becke--Johnson potential
\cite{Tran2009:PRL} parametrized to give the bandgap of $2.17$~eV.

The calculated band structure of phosphorene is shown in Fig.~\ref{fig:bs_dos}. We get a direct gap at the zone center.
Some DFT calculations \cite{Popovic_2015,rodin_strain_induced_2014,Ziletti_2015}
report a nearly indirect bandgap, with a somewhat displaced valence band maximum. 
The valence band in the vicinity of the $\Gamma$ point along $k_x$ is nearly dispersionless while it is very dispersive in the $k_y$ direction. 
Similar, but substantially smaller dispersion anisotropy is seen in the conduction band. 
Close to the $\Gamma$ point both, the valence and the conduction band have mainly $p_z$ orbital character,
the latter having a small admixture of $p_y$ orbitals \cite{SM}.
The next conduction band minimum appears at energy $50$~meV above the global conduction band minimum (in direction towards the X point) and consists mainly of $p_x$ and $p_y$ orbitals with an admixture of $d_{x^2-y^2}$ electrons.
The small distance of this band to the conduction band minimum is reflected as an increase of 
the slope in the carrier density $n(\varepsilon_{\rm F})$ shown in Fig.~\ref{fig:bs_dos}(e)
as a function of the Fermi level $\varepsilon_{\rm F}$. 
For the valence band the carrier density is a smooth quadratic function of $\varepsilon_{\rm F}$. Empirical
fits for the dependence $n(\varepsilon_F)$, which should be useful for interpreting experiments,
are given in the caption to Fig.~\ref{fig:bs_dos}.


\begin{figure}
\centering
\includegraphics[width=0.99\columnwidth]{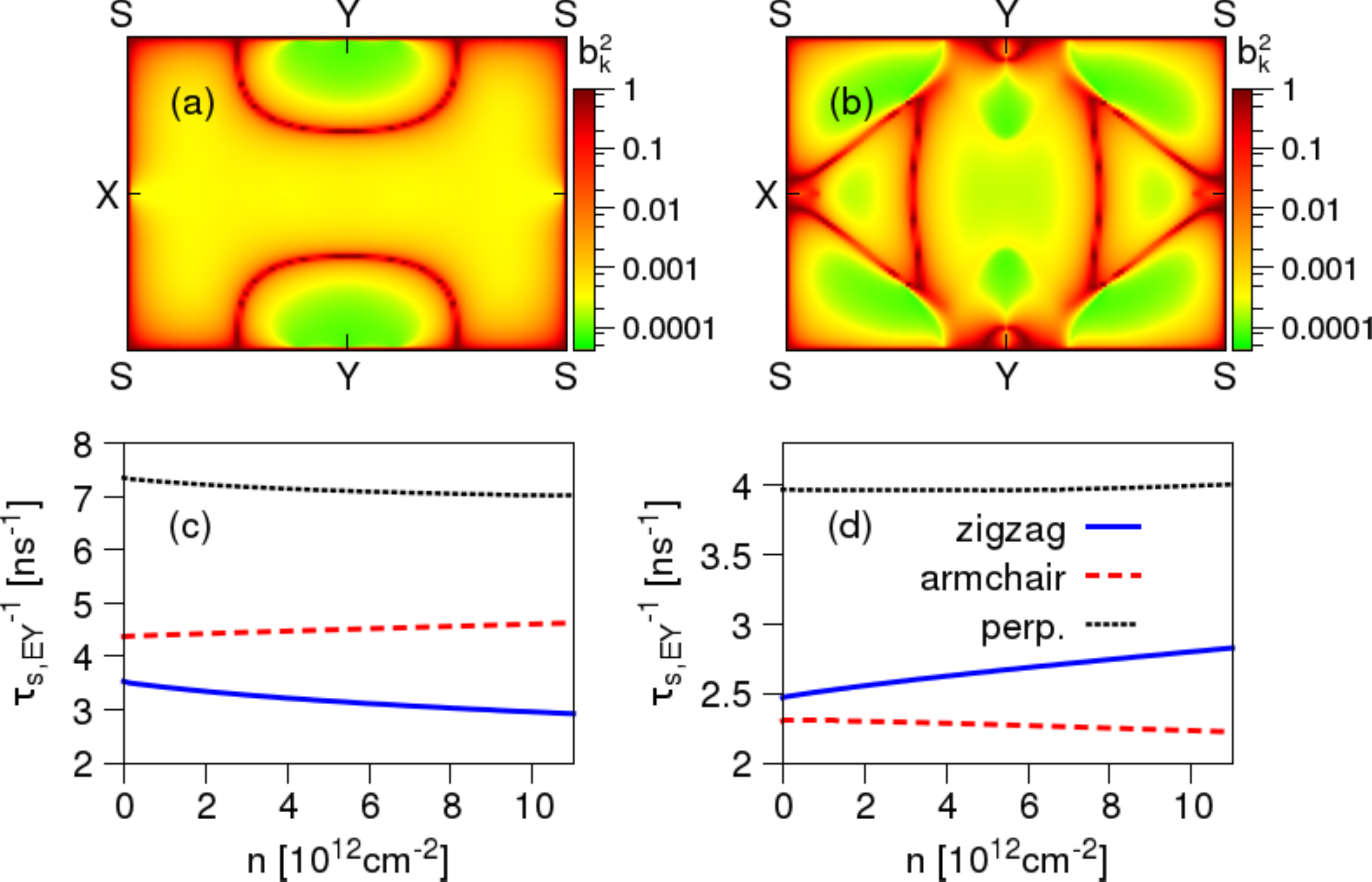}
\caption{\label{fig:4plot_ey} (Color online) 
Intrinsic spin-orbit coupling effects in phosphorene.
(a)~Momentum-resolved spin-mixing parameter $b^2_\mathbf{k}$ for the valence 
band and out-of-plane spin direction.
(b)~Same as in (a) but for the conduction band.
(c)~Elliott--Yafet spin relaxation rates for valence electrons, for indicated spin directions 
as a function of the carrier density. (d)~Same as in (c) but for the conduction band. Constant typical momentum
relaxation time $\tau_p=100$~fs is assumed.
}
\end{figure}

\paragraph{Intrinsic effects.} We first discuss the intrinsic SOC and the Elliott--Yafet spin relaxation.

The intrinsic SOC in phosphorene is relatively strong but does not modify
substantially the bandstructure close to the bandgap. 
The orbital degeneracy of the fourfold degenerate bands, see inset to Fig.~\ref{fig:bs_dos}(c), is split 
into two pairs of spin degenerate bands. 
The splitting is maximal at the S point, $17.5$~meV and $14$~meV for valence and conduction band respectively, gradually decreasing towards the time reversal points X and Y. The states at the Brillouin zone edges {\em sticking together} due to nonsymmorphicity of the $D_{2h}$ group \cite{Dresselhaus2007:book}. 

The important effect of the intrinsic SOC is the spin mixing, quantified by $b^2_\mathbf{k}$. 
In Fig. \ref{fig:4plot_ey}(a,b) we show the distribution of spin mixing parameter $b^2_\mathbf{k}$ in the first Brillouin zone of phosphorene for the spin quantization axis oriented perpendicular to the 2D plane. The other spin orientations are discussed in Ref.~\cite{SM}. 
For momenta corresponding to anticrossings and at the BZ edges (except the points X and Y for which $b^2_\mathbf{k}$ is zero) the values of $b^2_\mathbf{k}$ are close to $\frac{1}{2}$. 
The Bloch eigenstates here are fully spin-mixed, forming spin hot spots 
\cite{fabian_spin_1998, fabian_phonon-induced_1999}.
At the zone center $b^2_\mathbf{k}$ is about $10^{-4}$.
Perturbation theory gives that $b$ is roughly the ratio of the intrinsic spin-orbit coupling (order 10 meV) and the 
band gap (order 1 eV), thus $b \sim 0.01$, matches well to the calculated first-principles value of $b^2 \sim 10^{-4}$. 
For comparison, the $k\cdot p$ theory gives $b^2$ between $10^{-6}$ to $10^{-5}$ \cite{li_electrons_2014}.

\begin{figure}[h]
\centering
\includegraphics[width=0.99\columnwidth]{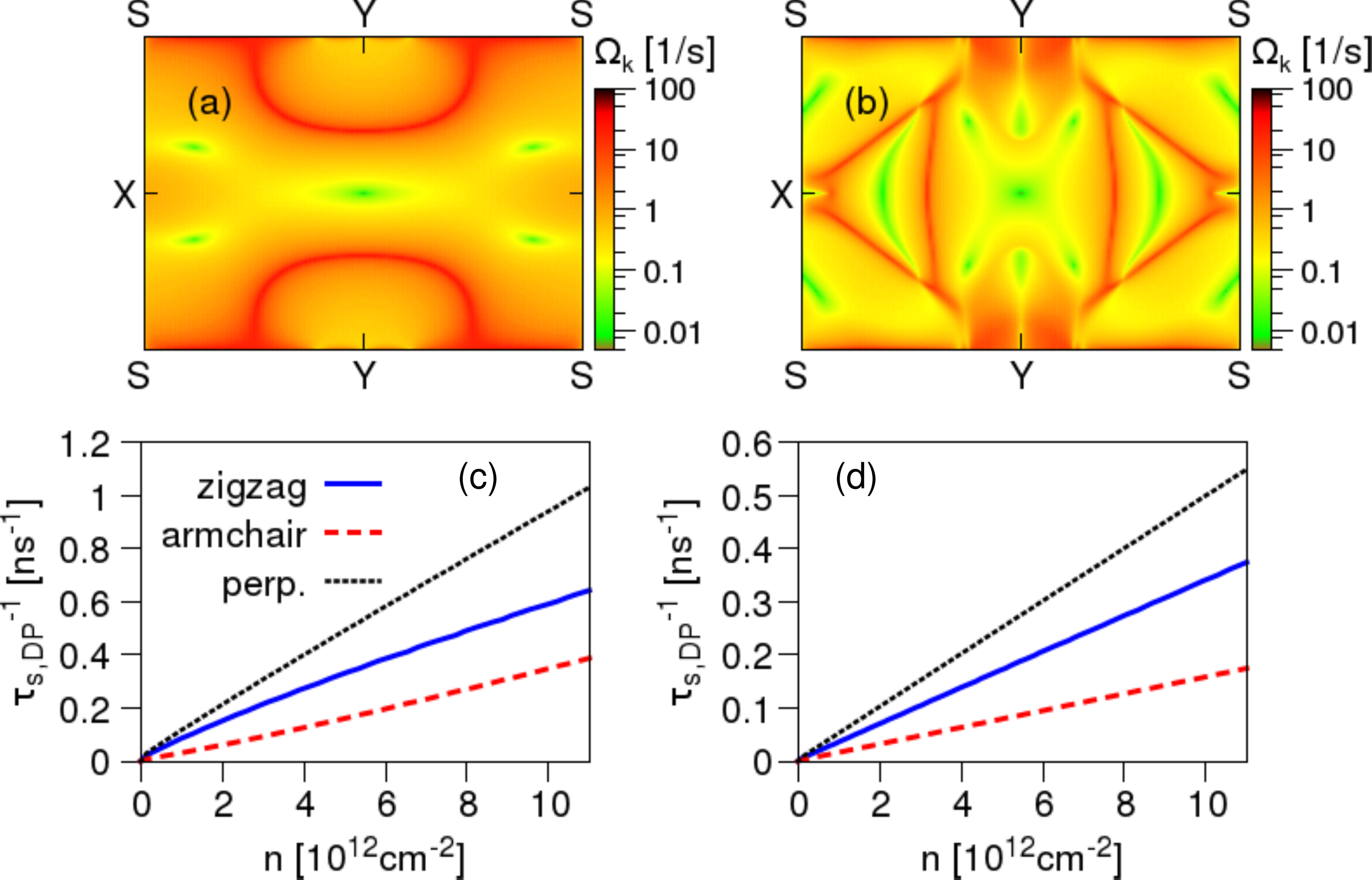}
\caption{\label{fig:DPrelax} (Color online)
Extrinsic spin-orbit coupling effects in phosphorene.
(a)~Spin-orbit field magnitude $\Omega_\mathbf{k}$ in the first Brillouin zone for the 
valence band and transverse electric field of $E=1$~V/nm.
(b)~Same as in (a) but for the conduction band.
Calculated D'yakonov--Perel' relaxation rates, assuming $\tau_p = 100$ fs, as a function of carrier density for (c)~valence band and (d)~for conduction band for indicated spin directions. 
For spins perpendicular to the phosphorene plane empirical fitting gives 
$\tau^{-1}_{\rm s, DP} (n)[\text{ns}^{-1}]\approx 0.093n$
for the valence band and $\tau^{-1}_{\rm s, DP}(n)[\text{ns}^{-1}]\approx 0.05n$ 
for the conduction band, where $n$ is in units of $10^{12}$~cm$^{-2}$.
}
\end{figure}

Knowing $b^2$ we now calculate the Elliott--Yafet spin relaxation rates using Eq.~(\ref{eq_ey}). For the 
momentum relaxation we take the typical experimental value of $\tau_p=100$~fs. 
The results can be easily rescaled for the actual experimental mobilities. Calculated $\tau_{\rm s, EY}^{-1}$ as a function of carrier density $n(\varepsilon_{\rm F})$, for valence and conduction bands and different spin quantization axes, are shown in Fig.~\ref{fig:4plot_ey}(c,d). 
The relaxation rates are almost independent of $n$, which follows $b^2$ since we use a constant momentum relaxation time. The monotonicity of $\tau_{\rm s,EY}^{-1}$ is then unambiguously determined by $b^2$. 
The spin relaxation rates of holes are greater than those of electrons. Most striking is the strong anisotropy.
The largest spin relaxation is for out-of-plane spins, which relax roughly twice as fast as the in-plane spins.
We predict the longest spin lifetimes for armchair-oriented spins in the conduction band, and zigzag-oriented
spins in the valence band. 
In the recent $k\cdot p$ theory \cite{li_electrons_2014}, the estimated ratio between the spin relaxation rate for out-of-plane 
to in-plane spins was $\sim 4$, which is an overestimation in view of our first-principles results, but is in a qualitative agreement. Similar anisotropies in Elliott--Yafet spin lifetimes
were also predicted for anisotropic bulk materials and thin metallic films \cite{zimmermann_2012,long_2013}.

\paragraph{Extrinsic effects.} In realistic situations phosphorene sits on a substrate or is studied in a gating 
electric field which breaks space inversion symmetry ($D_{2h}\longrightarrow C_{2v}$). 
An extrinsic Rashba spin-orbit field emerges, lifting the spin degeneracy, 
$\varepsilon_{\mathbf{k}\uparrow}\neq \varepsilon_{\mathbf{k}\downarrow}$, according to Eq.~(\ref{eq_ss}), 
except at time-reversal invariant points.
Emerged spin-orbit fields give rise to spin relaxation due to the D'yakonov--Perel' mechanism, 
which competes with the Elliott--Yafet spin-flip scattering. Here we model the symmetry breaking by 
applying a transverse electric field, all within the first-principles calculations, of 1 V/nm.  
In Fig.~\ref{fig:DPrelax}(a,b) we plot the spin-orbit field $\Omega_{\mathbf{k}}$ magnitude over the first Brillouin zone.
Similarly to $b^2_\mathbf{k}$, the values of $\Omega_{\mathbf{k}}$ are peaked at the bands anticrossings 
and at the BZ edges. 
At the time-reversal points the $\Omega_{\mathbf{k}}$ is zero. \\
We note, that for a bare PBE exchange-correlation functional \cite{Pedrew_1996} (band gap $E_g\approx 1$~eV), the Rashba spin-orbit coupling due to external electric fields exhibits a strong anisotropy in the valence as well as in the conduction band \cite{Popovic_2015}. 
Our calculations show that increasing the bandgap to the experimental value $\approx 2$~eV removes the anisotropy from the conduction band, while it is preserved for the valence band. 

To obtain the spin relaxation rates for the D'yakonov--Perel' mechanism, we 
resolve the coordinate components of the vector spin-orbit fields  $\Omega$ which
lie in the phosphorene plane: 
$\Omega_x$ along $x$ (zigzag) and $\Omega_y$ along $y$ (armchair) directions. 
We extract these components by fitting an effective $C_{2v}$ symmetric spin-orbit coupling Hamiltonian to the first-principles data \cite{SM}. 
In Fig.~\ref{fig:DPrelax}(c,d) we show the calculated spin relaxation rates using Eq.~(\ref{eq_dp}), assuming $E=1$~V/nm and $\tau_{p}=100$~fs. 
The spin lifetime is exceptionally long, of a few ns, and exceeds the lifetime from the Elliott--Yafet mechanism.
The relaxation rates for conduction electrons are twice smaller than for the valence electrons. 
With a growing electric field, the D'yakonov--Perel' mechanism becomes more significant.
For the valence band, it surpasses the Elliot--Yafet's for $n\gtrsim 6\cdot 10^{12}$~cm$^{-2}$, $n\gtrsim 3\cdot 10^{12}$~cm$^{-2}$ and $n\gtrsim 2\cdot 10^{12}$~cm$^{-2}$ for electric fields $E=3$~V/nm, $E=4$~V/nm and $E=5$~V/nm respectively. 
For the conduction band the transitions happen for slightly higher carrier densities.
For electric fields $E=5$~V/nm the D'yakonov--Perel' spin relaxation rates are $\approx 23$ times bigger than for $E=1$~V/nm.
Similarly to $b^2$, the spin-orbit field $\Omega^2$ reveals a strong anisotropy. 
As a result the in-plane spins relax about $1.5-3$ times slower than the out-of-plane ones.
		
In summary, we have studied intrinsic and extrinsic spin-orbit coupling and spin relaxation mechanisms in phosphorene. The Elliott--Yafet spin relaxation gives spin lifetimes less than nanoseconds for experimentally
relevant samples. The D'yakonov--Perel' mechanism matters at large electric fields. The lifetimes exhibit a large anisotropy for in-plane and out-of-plane spin orientations.

We acknowledge funding from DFG SPP 1538, SFB 689, NCN DEC-2013/11/B/ST3/00824 and the EU Seventh Framework Programme under Grant Agreement No. 604391 Graphene Flagship.

\bibliography{bib_lib1}

\begin{thebibliography}{41}
\bibitem [{\citenamefont {see Suplementary~information}()}]{SM}%
  \BibitemOpen
  \bibfield  {author} {\bibinfo {author} {\bibnamefont {see
  Suplementary~information}},\ }\href@noop {} {}\bibinfo {note} {For structural
  parameters, calculation of effective SOC Hamiltonian parameters, spin-orbit
  fields, spin mixing parameter $b^2$. The distribution of $b^2_{\mathbf{k}}$
  in the first Brillouiin zone and orbital decomposition of the band structure
  are also shown.}\BibitemShut {Stop}%
\end{thebibliography}

\newpage

\section*{Supplementary information}

\subsubsection*{Structural parameters}

\begin{figure}[h]
\centering
\includegraphics[width=0.6\columnwidth]{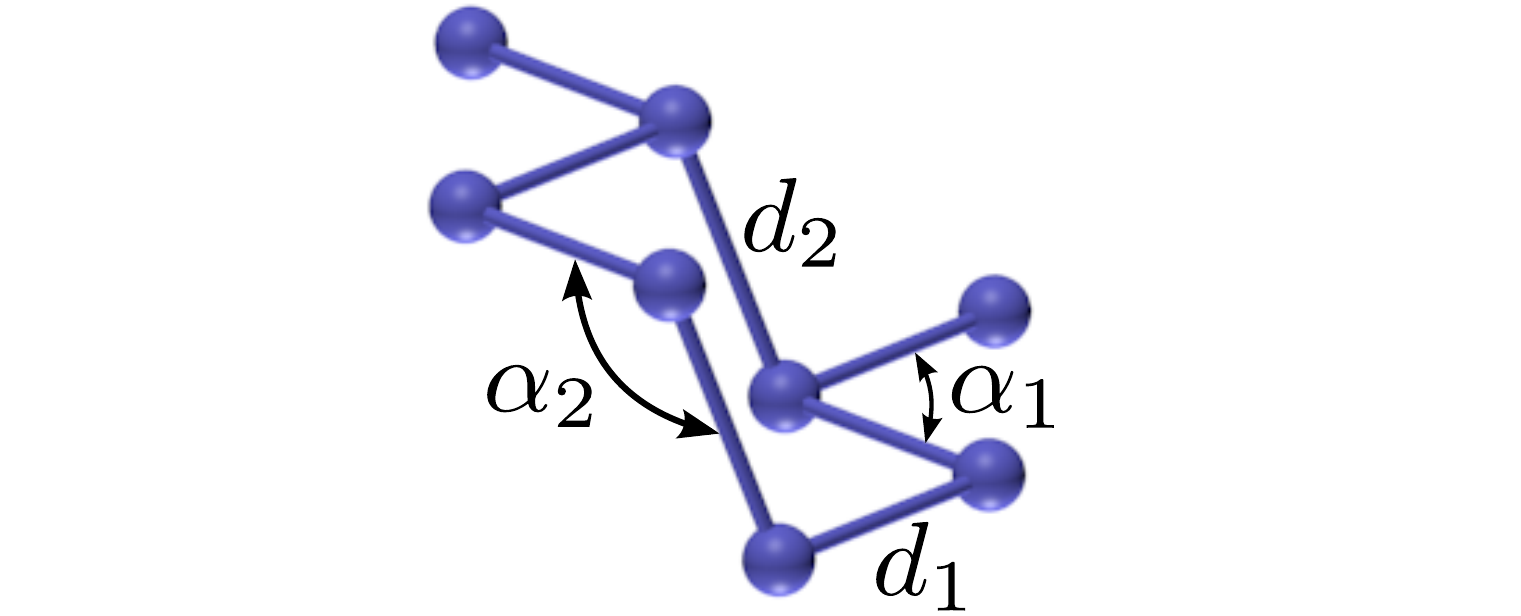}
\caption{\label{fig:bonds} (Color online) Sketch of the structural parameters summarized in Table \ref{tab:table1}.
}
\end{figure}

Structure parameters for phosphorene obtained by structural relaxation \cite{QE-2009} and experimental values for black phosphorus \cite{brown_refinement_1965} are given in Tab.~\ref{tab:table1}. The lattice constants, interlayer bond distances and bond angles are sketched in Fig.~\ref{fig:bonds}.

\begin{table}[h]
\caption{\label{tab:table1}%
Structural parameters for relaxed phosphorene and experimental data for bulk black phosphorus. }
\begin{ruledtabular}
\begin{tabular}{ccc}
& 2D relaxed & bulk exp. \citep{brown_refinement_1965}\\ 
\hline
a &  3.2986	\AA & 3.3136 \\
b & 4.6201 \AA & 4.376 \\
$d_1$ & 2.2223 \AA & 2.224 \\
$d_2$ & 2.2601 \AA & 2.244 \\
$\alpha_1$& 95.833 & 96.34 \\
$\alpha_2$& 104.085 & 102.09 \\
\end{tabular}
\end{ruledtabular}
\end{table}

\subsubsection*{Spin--mixing parameters}

\begin{figure}[h]
\centering
\includegraphics[width=0.98\columnwidth]{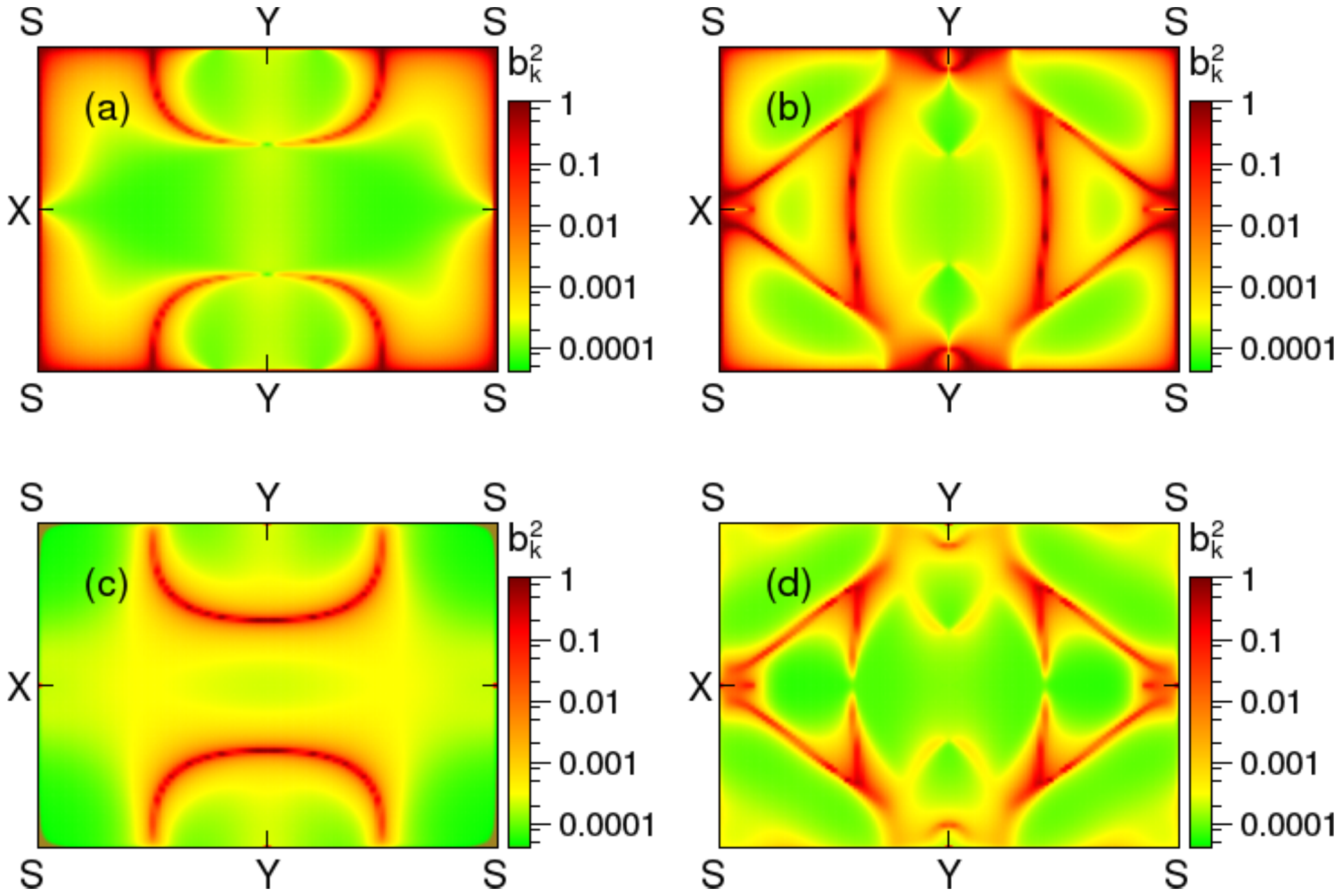}
\caption{\label{fig:b2kmap_suppl} (Color online) 
Momentum resolved spin mixing parameter $b^2_\mathbf{k}$ for in-plane spin orientation in phosphorene. The calculated mixing parameter for (a)~valence band and zigzag spin orientation, (b)~conduction band and spin along zigzag direction, (c)~valence band and spin pointing along armchair direction and (d)~conduction band with spin along armchair direction. The spin hot-spots ($b^2_k$ close to 0.5) are identified at the zone edges and at accidental bands anticrossings. The values of $b^2_\mathbf{k}$ at the zone center are shown in Fig. \ref{fig:b2_anisotropy} for zero carrier density. 	 
}
\end{figure}

\begin{figure}[h]
\centering
\includegraphics[width=0.99\columnwidth]{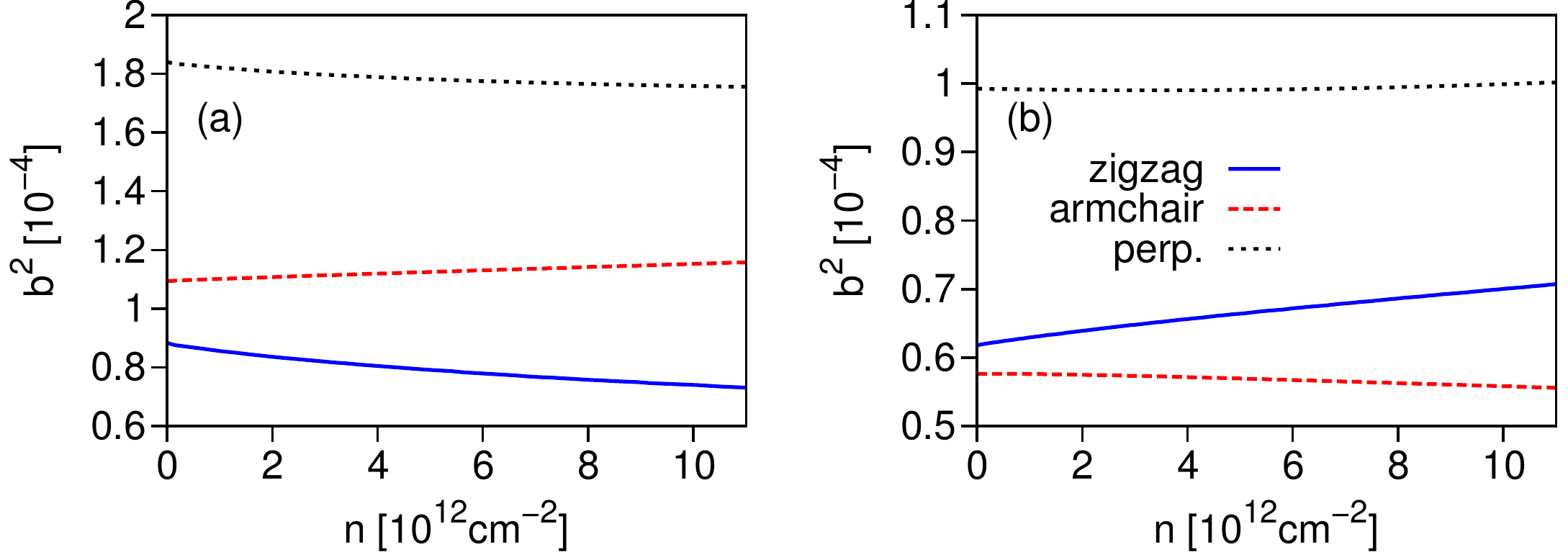}
\caption{\label{fig:b2_anisotropy} (Color online) 
Fermi contour averaged spin mixing parameter $b^2$ versus carrier density $n$ in units $10^{12}$~cm$^{-2}$, for different spin quantization axes for (a)~valence band, and (b)~conduction band. The values of $b^2$ for spins oriented perpendicularly to the phosphorene plane are approximately twice bigger than for in-plane ones.
Weaker intrinsic SOC in the conduction band gives roughly twice smaller values of $b^2$ that for electrons than for holes. 
}
\end{figure}

In Fig.~\ref{fig:b2kmap_suppl} we plot the distribution of the spin-mixing parameter $b^2_\mathbf{k}$ in the first Brillouin zone for phosphorene and spin quantization axis along the zigzag [Fig. \ref{fig:b2kmap_suppl}(a,b)] and the armchair direction [Fig. \ref{fig:b2kmap_suppl}(c,d)]. 
The differences between the two spin quantization axes are remarkable. 
For the zigzag spin quantization axis the spin hot spots are located both at the bands anticrossings and at the BZ edges, whereas for the armchair one the hot regions are only at the anticrossings. 
Close to the zone center, around the maximum of the valence band and the minimum of the conduction band, the differences in $b^2_\mathbf{k}$ are not so evident. The corresponding values of $b^2$, calculated versus carrier density, are similar for the zigzag and armchair spin quantization axes, as shown in Fig. \ref{fig:b2_anisotropy}. 
The parameter $b^2$ takes the largest values for the spins oriented out-of-plane, giving the shortest spin lifetimes. 
The spin quantization axis with the minimal values of $b^2$ (and the longest spin lifetime) is along the zigzag and the armchair edges of phosphorene for the valence and the conduction band respectively. The ratio between maximal and minimal $b^2$ is roughly 2.25 for holes and 1.8 for electrons. 
 
\subsubsection*{ Effective spin--orbit Hamiltonian and spin--orbit fields}
The extrinsic SOC effects close to the $\Gamma$ point can be described by the effective
spin-orbit Hamiltonian preserving $C_{2v}$ symmetry,
$H_{\rm soc} = \alpha \left( \sigma_y k_x -\sigma_x k_y \right)+ \gamma\left( \sigma_y k_x +\sigma_x k_y\right)$,
where $\alpha$ and $\gamma$ are the SOC parameters, akin for Rashba and Dresselhaus SOC 
in semiconductor heterostructures \cite{Bychkov1984:JETP}, $k_x$ and $k_y$ are the 
components of the wave vector, and $\sigma_x$ and $\sigma_y$ are the Pauli matrices.
Diagonalizing $H_{\rm soc}$ we can express the energy spin splitting $\Delta\varepsilon(\mathbf{k})=2\sqrt{(\gamma +\alpha)^2k_x^2+(\gamma-\alpha)^2k_y^2}$ 
along principal axes in the form $\Delta\varepsilon=2|\alpha +\gamma| k_x$ and 
$\Delta\varepsilon=2|\alpha -\gamma| k_y$. The calculated spin-orbit coupling parameters $\alpha$ and $\gamma$ are listed in Tab.~\ref{tab:table2} for several experimentally relevant values of transverse electric field. 
\begin{table}
\caption{\label{tab:table2}%
Spin-orbit coupling parameters extracted from first-principles calculations for different values of the electric field. }
\begin{ruledtabular}
\begin{tabular}{ccccc}
 &\multicolumn{2}{c}{Valence band}& \multicolumn{2}{c}{Conduction band} \\
E [V/nm]& $\gamma$~[meV\AA]&$\alpha$~[meV\AA]& $\gamma$~[meV\AA]&$\alpha$~[meV\AA]\\
\hline
1.0 & 0.46 & -0.23 & 0.29 & -0.06 \\
2.0 & 0.91 & -0.48 & 0.56 & -0.11\\
3.0 & 1.36 & -0.72 & 0.84 & -0.17 \\
5.0 & 2.21 & -1.13 & 1.43 & -0.29
\end{tabular}
\end{ruledtabular}
\end{table}

\begin{figure}[h]
\centering
\includegraphics[width=0.99\columnwidth]{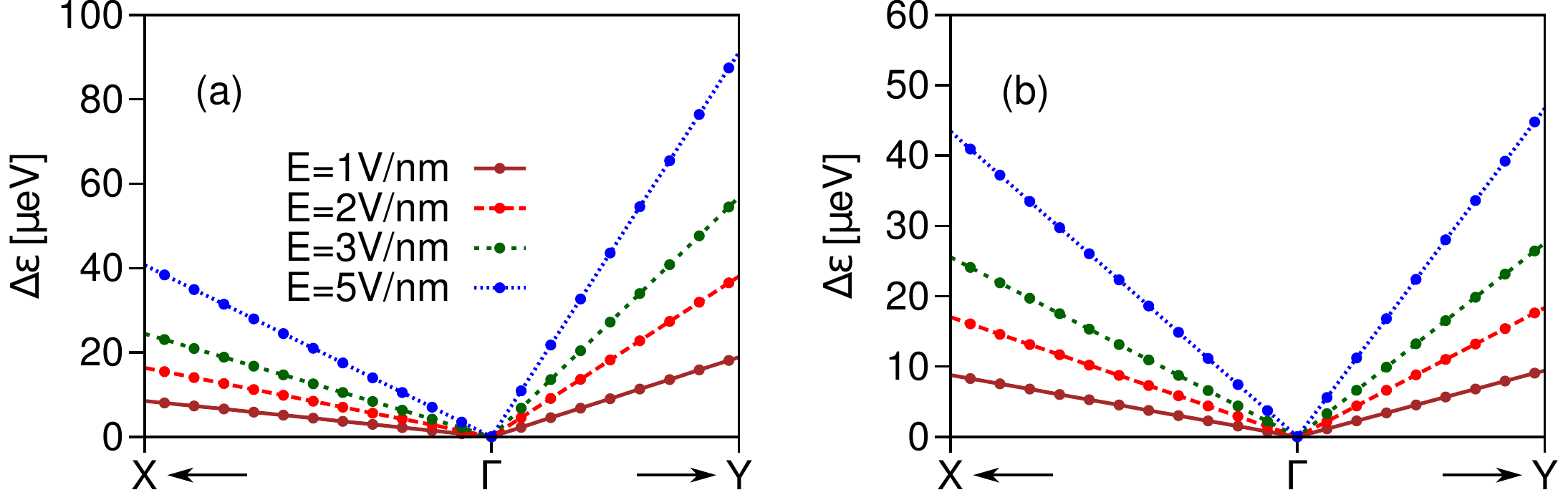}
\caption{\label{fig:splitting_vs_ef} (Color online) Energy spin splitting in vicinity of the zone center for several values of the external electric field $E$. The range of $k$ values covers 2$\%$ of $\Gamma$--X and $\Gamma$--Y widths. 
Energy spin splitting $\Delta\varepsilon$ for (a)~valence bands and (b)~for conduction band.
}
\end{figure}

In Fig.~\ref{fig:splitting_vs_ef} we plot the spin splitting $\Delta\varepsilon$ in the vicinity of the 
$\Gamma$ point for the valence and conduction bands for several values of the transverse electric field. 
For the valence band the splittings along $\Gamma$--X are twice smaller that for $\Gamma$--Y path. The anisotropy is much less pronounced in the conduction band. 

\begin{figure}[h]
\centering
\includegraphics[width=0.99\columnwidth]{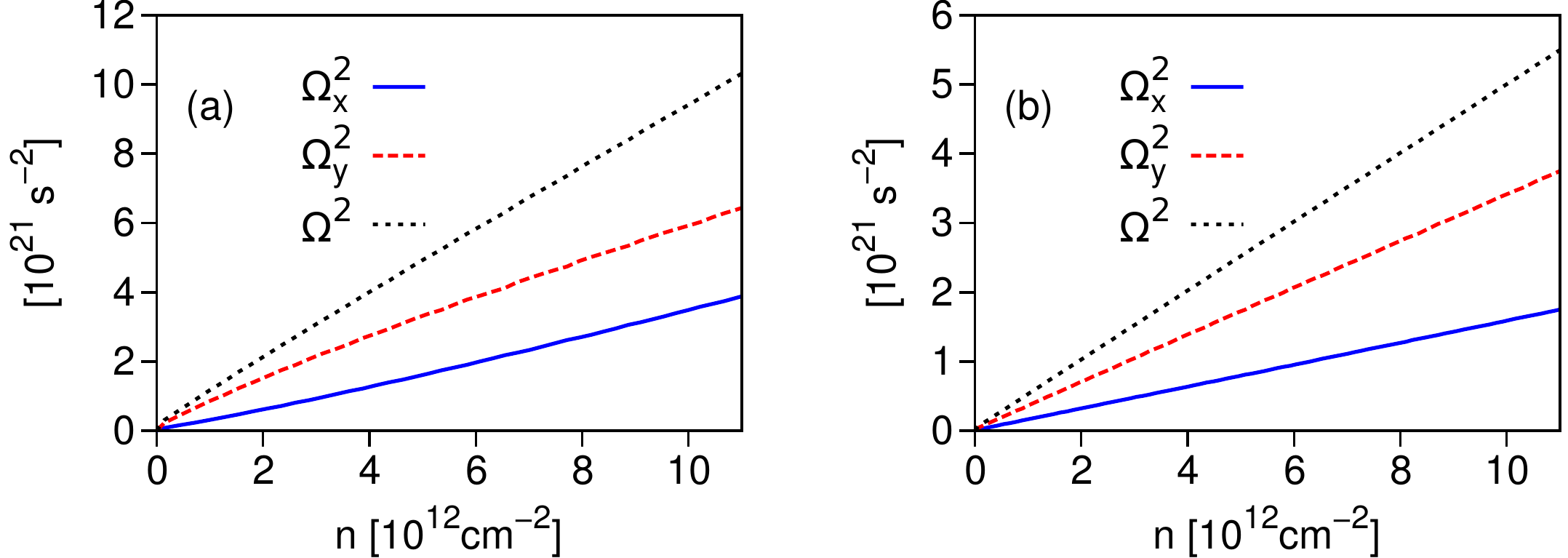}
\caption{\label{fig:Omega_anisotropy} (Color online) 
Fermi contour averaged squared spin-orbit fields $\Omega^2$ (black dotted line),  $\Omega^2_x$ (solid blue line) and $\Omega^2_y$ (dashed red line) versus carrier density for (a)~ valence band, and (b)~conduction band and transverse electric field $E=1$V/nm.
}
\end{figure}

Having the parameters $\alpha$ and $\gamma$ of the effective Hamiltonian $H_{\rm soc}$ one can calculate averages of squared spin--orbit field components: $\Omega^2_x$, $\Omega^2_y$, and $\Omega^2=\Omega^2_x+\Omega^2_y$
\begin{eqnarray}
\Omega_x^2 &=\frac{4(\alpha -\gamma)^2}{\hbar^2} \langle k_y^2 \rangle,\\
\Omega_y^2 &=\frac{4(\alpha +\gamma)^2}{\hbar^2}\langle k_x^2\rangle, 
\end{eqnarray}
where $k_x$ ($k_y$) is the momentum of carriers moving along zigzag (armchair) direction of phosphorene; the average is taken over the Fermi contour.
The spin-orbit fields are plotted in Fig.~\ref{fig:Omega_anisotropy} as a function of the carrier doping. 
The calculated values of $\Omega^2$ for the valence band are roughly twice as large as for the conduction band.
A significant anisotropy with respect to main crystal axes is observed giving the ratio between the maximal ($\Omega^2$) and the minimal ($\Omega^2_x$) values about  2.5 for the valence and 3 for the conduction band. 

\begin{figure}
\centering
\includegraphics[width=0.99\columnwidth]{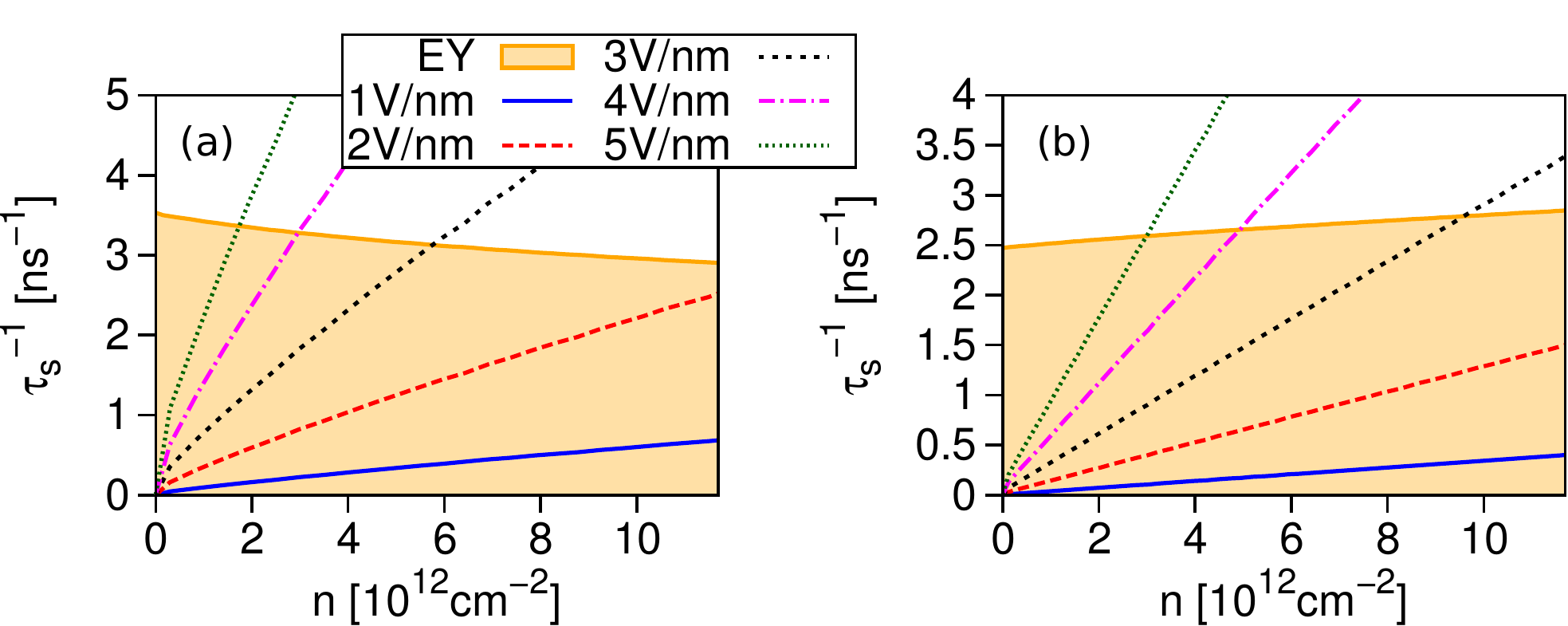}
\caption{\label{fig:ey_vs_dp} (Color online) Comparison of the Elliott--Yafet (shaded) and D'yakonov--Perel' spin relaxation times versus carrier density for several values of the electric field and zigzag spin quantization axis; a) valence band, b) conduction band. The momentum relaxation time $\tau_p=100$~fs is assumed.
}
\end{figure}

\subsubsection*{ Elliott--Yafet versus D'yakonov--Perel'}
We compare the spin scattering rates from the Elliott--Yafet and the D'yakonov--Perel' mechanisms for spins pointing along zigzag direction in Fig.~\ref{fig:ey_vs_dp}. For electric fields $E\le 2$~V/nm the Elliott--Yafet dominates spin scattering of electrons and holes. 

\begin{figure}
\centering
\includegraphics[width=0.99\columnwidth]{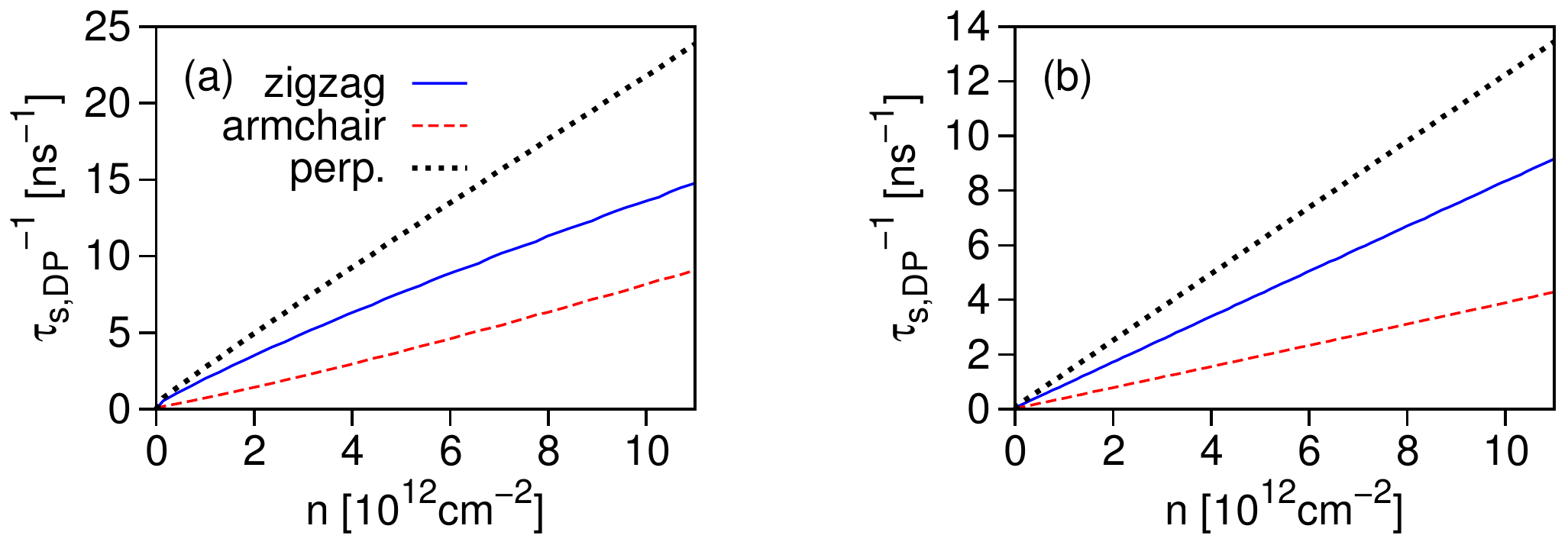}
\caption{\label{fig:dp_relaxrate_5Vnm} (Color online) Calculated D'yakonov--Perel' relaxation rates versus carrier density (in units $10^{12}$~cm$^{-2}$) for electric field $E=5$~V/nm and for indicated spin orientations. a) valence band, b) conduction band. Constant typical momentum relaxation time $\tau_p=100$~fs is assumed. }
\end{figure}

With increasing electric field the contribution of the D'yakonov--Perel' mechanism to spin relaxation increases and depends on carrier type and density. 
For $E=5$~V/nm it overtakes the Elliott--Yafet's for carrier densities $n\ge 2 \cdot 10^{12}$~cm$^{-2}$ for holes and  $n\ge 3 \cdot 10^{12}$~cm$^{-2}$ for electrons. 
Increasing carrier concentration the relaxation rates grow up to 25~ns$^{-1}$, see Fig. \ref{fig:dp_relaxrate_5Vnm}(a,b), and the spin lifetime becomes mainly limited by the D'yakonov--Perel' mechanism for all spin directions and carrier types.

\subsubsection*{Orbital resolved bandstructure}
\begin{figure*}[h!]
\centering
\includegraphics[width=\textwidth]{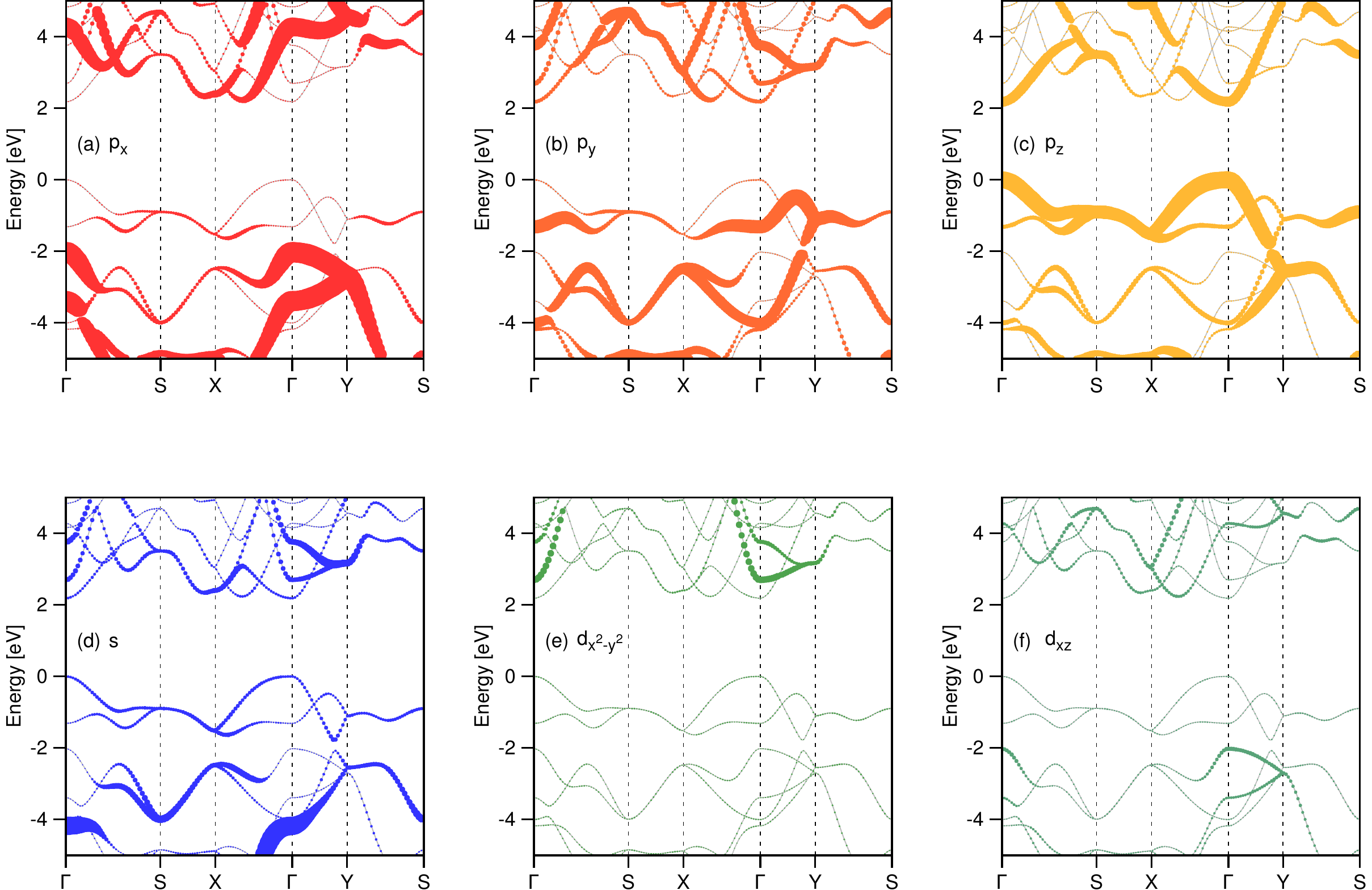}
\caption{\label{fig:band_character} (Color online) Calculated orbital decomposition of the phosphorene band structure. Corresponding orbital contributions to the states are proportional to the symbol radii for
(a)~$p_x$, (b)~$p_y$,(c)~$p_z$, (d)~$s$, (e)~$d_{x^2-y^2}$, (f)~$d_{xy}$ orbitals.
}
\end{figure*}

In figure~\ref{fig:band_character} we show calculated $s$, $p$ and $d$ orbital resolved band structure plots along the high symmetry lines in the first Brillouin zone.
The valence and conduction band edges near the zone center are formed by $p_z$ orbitals.
The first valence band preserves its $p_z$ orbital character within bandwidth of about 2~eV and then further mix with other bands of $p_x$ and $p_y$ orbital character.
In detail we see that the first valence band along all high symmetry lines is almost purely of $p_z$ character with a small contribution from the $s$ electrons, see Fig.~\ref{fig:band_character}(c,d). 
Exception is along the $\Gamma$-$Y$ line when the $p_y$ band rises in energy with increasing momentum $k$. At the $Y$ point the two $p_y$ bands merge and {\em stick together} at the zone edge towards the $S$ point. Along this way the stuck bands change their character from $p_y$ to $p_z$.
At the energies about $2$~eV below the Fermi level, a contribution from $d_{xy}$ orbital character is found which is an admixture to the $p_x$ band, Fig.~\ref{fig:band_character}(a) and (f).

For the first conduction band the situation is similar to the first valence band. An important difference is noted along the $\Gamma$-$X$ line where another parabolic-like band of mixed $p_x$ and $p_y$ character crosses the $p_z$ band. The energy offset of the band minimum is about 50~meV above the conduction band edge at the zone center. Presence of this band influences charge carrier concentration as shown in Fig.~2(e) in the main text.
Contrary to the valence band a contribution from the $d$ electrons in the conduction band manifold is found at much lower energies. At about 0.5~eV from the conduction band edge there is a significant contribution from the $d_{x^2-y^2}$ electrons to the second conduction band, see Fig.~\ref{fig:band_character}.

\end{document}